\documentclass[reprint,
 bibnotes,
 amsmath,amssymb,
 aps,
prx,
floatfix,
]{revtex4-2}

\usepackage{soul}
\usepackage{graphicx}% Include figure files
\usepackage{dcolumn}% Align table columns on decimal point
\usepackage{bm}% bold math
\usepackage{xcolor}
\usepackage{amssymb}

\usepackage{subfigure}

\usepackage{mathptmx}

\usepackage{amsmath}

\usepackage{natbib}
\begin{document}

\title{Influence of imperfections on tunneling rate in $\delta$-layer junctions}

\author{Juan P. Mendez}
\email{jpmende@sandia.gov}
\author{Shashank Misra}
\author{Denis Mamaluy}
\email{mamaluy@sandia.gov}

\affiliation{Sandia National Laboratories, Albuquerque, New Mexico, 87123}%

\date{\today}

\begin{abstract}

The atomically precise placement of dopants in semiconductors using scanning tunneling microscopes has been used to create planar dopant-based devices, enabling the exploration of novel classical or quantum computing concepts, which often require precise control over tunneling rates in their operation. While the geometry of the dopants can be defined to sub-nanometer precision, imperfections can still play a significant role in determining the tunneling rates. Here, we investigate the influence of different imperfections in phosphorous $\delta$-layer tunnel junctions in silicon: variations of $\delta$-layer thickness and tunnel gap width, interface roughness, and charged impurities. It is found that while most of the imperfections moderately affect the tunneling rate, a single charged impurity in the tunnel gap can alter the tunneling rate by more than an order of magnitude, even for relatively large tunnel gaps. Moreover, it is also revealed that the tunneling rate strongly depends on the electrical charge sign of the impurity.

\end{abstract}

\maketitle

\section{Introduction}\label{sec:introduction}

\begin{figure}[ht!]
  \centering
  \includegraphics[width=0.7\linewidth]{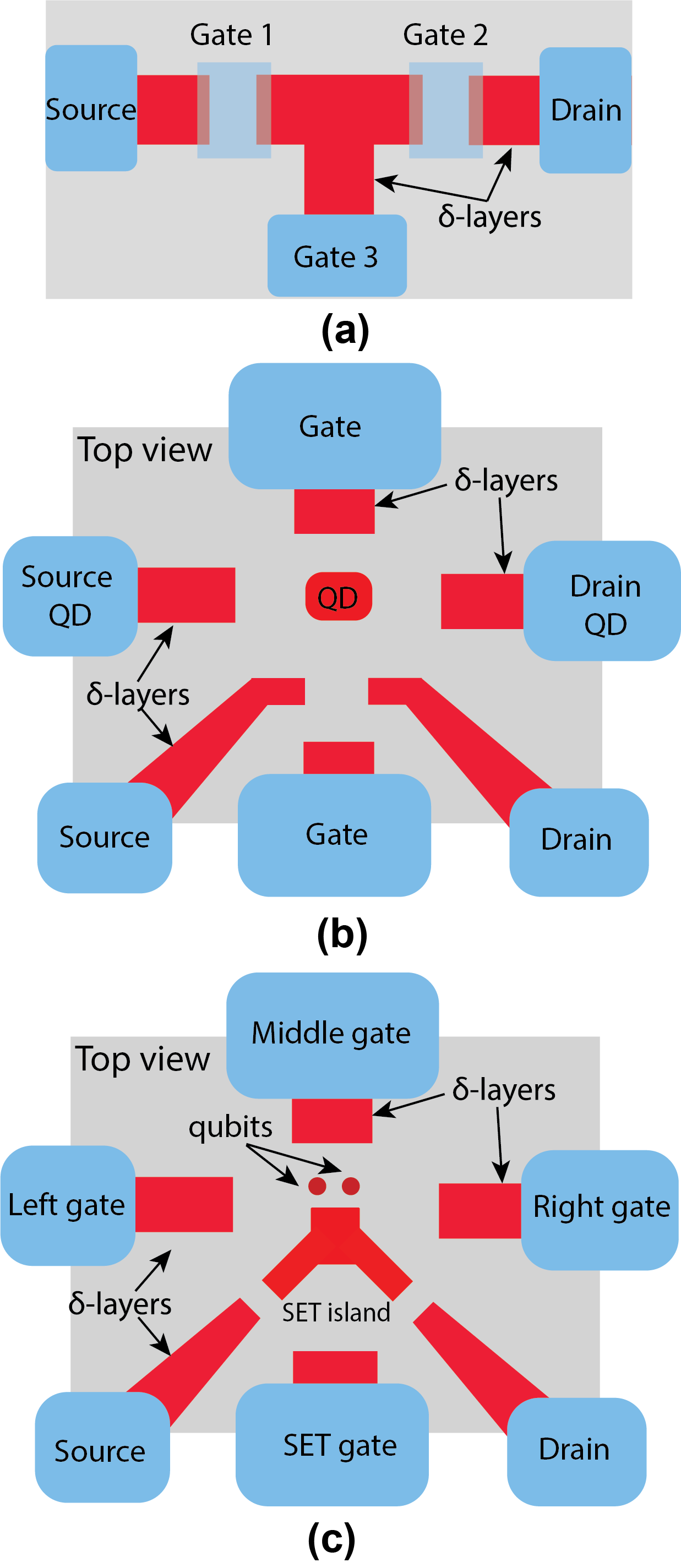}
  \caption{\textbf{Examples of atomic-precision advanced manufacturing nanodevices.} \textbf{a} $\delta$-layer tunnel junction field-effect transistor (FET) introduced by \cite{Donnelly:2021}. \textbf{b} A $\delta$-layer tunnel junction is used for detecting electrons in the quantum dot (QD) in \cite{House:2014}. \textbf{c} Two-qubit donor spin device from \cite{He:2019, Kranz:2020}. }
  \label{fig:example of APAM devices}
\end{figure}

Atomic precision advanced manufacturing (APAM) can be used to create 2D doped regions (known as $\delta$-layers) in silicon that simultaneously have single-atom precision \cite{Wilson:2006,Warschkow:2016,Fuechsle:2012,Ivie:2021b,Wyrick:2022,Campbell:2022} and very high conductivity \cite{Goh:2006,Weber:2012,McKibbin:2013, Keizer:2015, vskerevn:2020,Dwyer:2021}. APAM has application for exploring basic principles of novel electronic devices, including nano-scale diodes and transistors for classical computing and sensing systems \cite{Mahapatra:2011, House:2014, vskerevn:2020,Donnelly:2021} (see e.g.  Fig.~\ref{fig:example of APAM devices} \textbf{a} and \textbf{b}). Primarily, however, this technology is used to explore dopant-based qubits in silicon, with recent advancements in understand exchange-based 2-qubit operations \cite{He:2019}, the limits to qubit fidelity from environmental noise \cite{Kranz:2020}, the advantages to leveraging the number of dopants as a degree of freedom \cite{Krauth:2022,Fricke:2021}, and the exploration of many body \cite{Wang:2022} and topological \cite {Kiczynski:2022} effects in dopant chains (see e.g. Fig.~\ref{fig:example of APAM devices} \textbf{c}). In principle, atomically precise fabrication imbues the kind of control required by these applications, which have a high sensitivity to tunnel rates.

However, in reality, APAM involves tradeoffs between a number of defect mechanisms whose impact on tunnel rates have not been systematically studied, and this work pursues. A general processing tradeoff exists where the point defect density can be reduced by increasing the various processing temperatures, at the expense of worse dopant placement uncertainty from activating dopant diffusion \cite{Keizer:2015}. More specifically, there is a well-known intrinsic stochasticity from the underlying chemistry resulting in a dopant placement uncertainty of $\pm 0.3$~nm \cite{Wilson:2006,Ivie:2021b}. Moreover, after dopant incorporation, $\delta$-layer devices must be capped  with silicon at moderate temperature to protect them, but adatom-mediated diffusion can lead to a loss of out-of-plane sharpness that is on the order of 1 nm \cite{Hagmann:2018}. In contrast, the low temperature capping layer growth also leads to charged point defects at a density of $\sim 1$ defect in a ($10$~nm)$^3$ volume \cite{Anderson:2020}. 
Determining which of these disorder mechanisms is most likely to create large variations in tunneling rates in a simpler tunnel junction geometry will help inform how to navigate these tradeoffs, and lays the groundwork to analyze the more complicated case of qubits in the future.

Electron tunneling is understood in different terms at different length scales – ranging from semi-empirical descriptions of resonant tunnel diodes in terms of effective barrier heights \cite{Sze:2006} to atomistic hopping through single molecules in break junctions \cite{Gehring:2019}. Challenges arise in problems where the macro-scale description of tunneling in terms of band structure cannot account for atomic-scale details by simply renormalizing parameters. A direct quantum-mechanical way to investigate tunneling requires an open-system charge transport treatment \cite{Ferry_2022}. In this work we will employ an efficient, charge self-consistent, quantum transport implementation of open-system Non-Equilibrium Green Function (NEGF) formalism, known as the Contact Block Reduction (CBR) method \cite{Mamaluy:2003,Mamaluy_2004,Sabathil_2004,Mamaluy:2005,Khan:2007,Khan_2008,Gao:2014,Mendez:2021}. We combine it with an effective mass description for free electrons, shown to be in a very good agreement with tight-binding models for Si nanowires with sizes down to 3~nm \cite{Wang:2005,Neophytou:2008} and P $\delta$-layer tunnel junctions in Si \cite{Supplementary_material}, to assess the effect of imperfections on the tunneling rate for phosphorous $\delta$-layer tunnel junctions in silicon (Si: P $\delta$-layer tunnel junction). The considered imperfections in this work include variations of the $\delta$-layer thickness, small variations of the tunnel junction gap length, roughness in the edges of the $\delta$-layers, and the presence of charged impurities in the intrinsic gap. 

\begin{figure}
  \centering
  \includegraphics[width=0.8\linewidth]{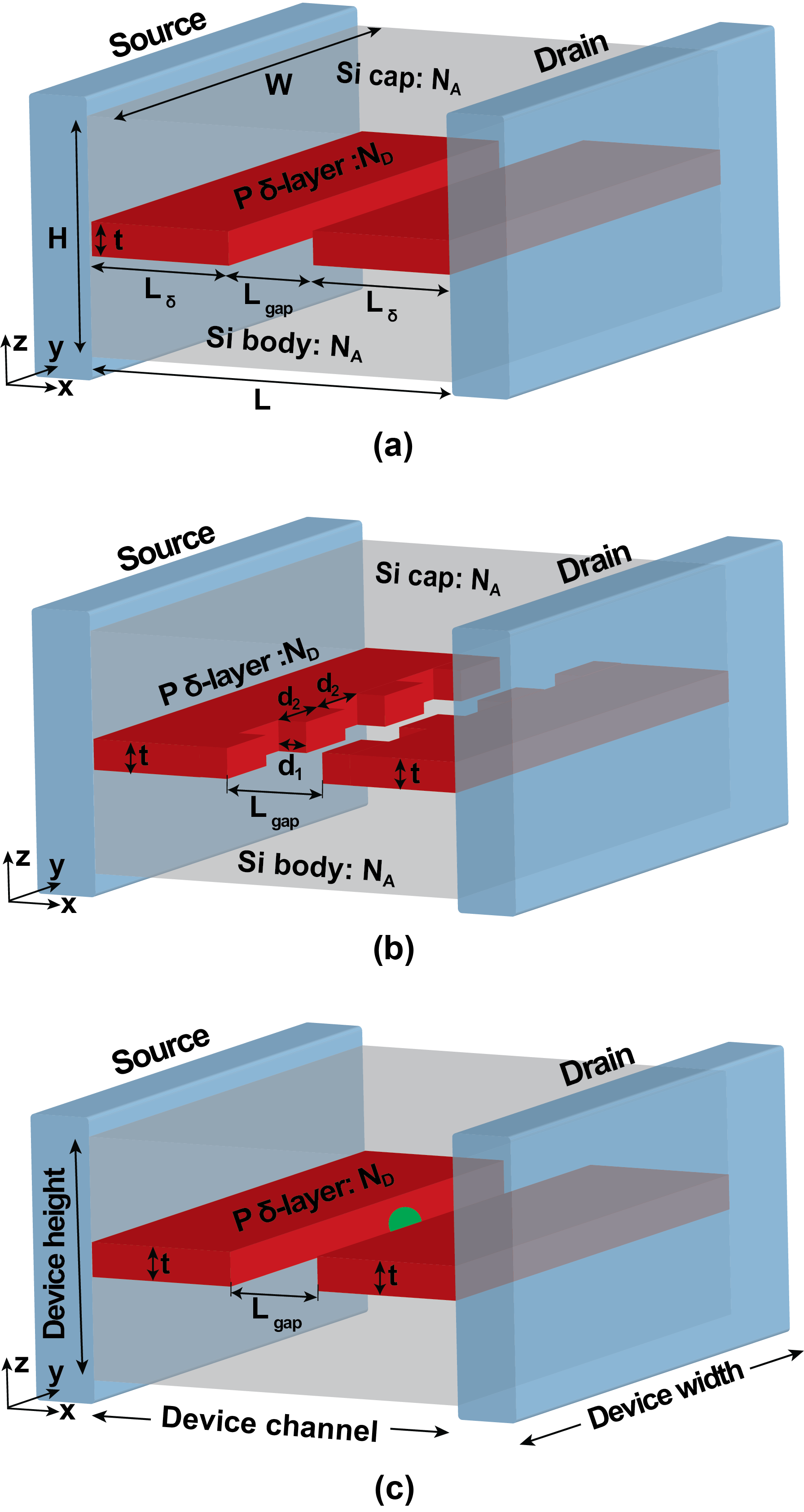}
  \caption{\textbf{Si:P $\delta$-layer tunnel junction (TJ) devices.} \textbf{a} Ideal device, which consists of a semi-infinite source and drain, in contact with the channel. The channel is composed of a lightly doped Si body and Si cap and a very thin, highly P doped-layer with an intrinsic gap of length $L_{gap}$. \textbf{b} Device with roughness in $\delta$-layer edges; The edge roughness is modeled as periodic protrusions of size $d_1 \times d_2 \times t$ with a periodicity of $2\times d_2$. \textbf{c} Device with presence of a charged impurity, either p-type or n-type, in the center of the intrinsic tunnel gap; The charged impurity is represented as a green sphere in the figure.}
  \label{fig:TJ_model}
\end{figure}

\section{Methodology}\label{sec:method}

To explore the impact of these defect mechanisms, we adopt a structure of a $\delta$-layer tunnel junction, which consist of two highly-conductive $\delta$-layers separated by an intrinsic semiconductor gap. In the open-system NEGF framework, the computational device consists of a semi-infinite source and drain, in contact with the channel of length $L$, which is composed of a lightly doped Si body and Si cap and two very thin, highly P-doped layers (referred to as left and right $\delta$-layers) separated by intrinsic gap of length $L_{gap}$, as shown in Fig.~\ref{fig:TJ_model} \textbf{a}. The channel length is chosen to be $L = 30~\text{nm} + L_{gap}$  to avoid the boundary effect between the source and drain contacts, the device height is $H=8$~nm and the total device width is chosen to be $W=15$~nm, with an effective width of 12~nm for the $\delta$-layer, to avoid size quantization effects on the conductive properties of $\delta$-layer systems \cite{Mendez_CS:2022}. We consider three different $\delta$-layer thicknesses: $t=0.2$~nm to approximate the true mono-atomic $\delta$-layer, an intermediate $t=1.0$~nm and the "thick" $\delta$-layer of $t=5.0$~nm. The sheet doping density of $\delta$-layer is $N_D=1.0 \times 10^{14}$~cm$^{-2}$ ($N_D^{(2D)} = t \times N_D^{(3D)}$) and the doping densities in the Si cap and Si body are $N_A=5.0 \times 10^{17}$~cm$^{-3}$ in all simulations. In work \cite{Lee:2011}, it was found that the effect of dopant (dis)order in the $\delta$-layers is negligible, thus modeling the $\delta$-layer as a continuum step-like doping profile is a good approximation. Furthermore, all simulations are carried out at the cryogenic temperature of $4$K, for which we can neglect inelastic scatterings \cite{Goh:2006,Mazzola:2014}. 

The simulations in this work are conducted using the open-system charge self-consistent Non-Equilibrium Green Function (NEGF) Keldysh formalism \cite{Keldysh:1965}, together with the Contact Block Reduction (CBR) method \cite{Mamaluy:2003,Mamaluy_2004,Sabathil_2004,Mamaluy:2005,Khan:2007,Khan_2008,Gao:2014,Mendez:2021} and the effective mass theory. The CBR method allows a very efficient calculation of the density matrix, transmission function, etc. of an arbitrarily shaped, multiterminal two- or three-dimensional open device within the NEGF formalism and scales linearly $O(N)$ with the system size $N$. As validation, in our previous works \cite{Mendez:2020,Mamaluy:2021}, we demonstrated a very good agreement with experimental electrical measurements for Si: P $\delta$-layer systems \cite{Goh:2006,Goh:2009,Reusch:2008,McKibbin:2014}, proving a excellent reliability of this framework to investigate $\delta$-layer systems. Similarly, our published results in \cite{Mendez_CS:2022}, without fitting parameters, agree remarkably well with the most recent experimental data for tunnel junctions in these systems\cite{Donnelly:2023}, as exhibited in the supplementary material\cite{Supplementary_material}.

Within this framework, we solve self-consistently the open-system effective mass Schr\"{o}dinger equation and the non-linear Poisson equation \cite{Mamaluy:2003,Mamaluy:2005,Gao:2014}. We employ a single-band effective mass approximation with a valley degeneracy of $d_{val}=6$. For the charge self-consistent solution of the non-linear Poisson equation we use a combination of the predictor-corrector approach and Anderson mixing scheme \cite{Khan:2007,Gao:2014}. First, the Schr\"{o}dinger equation is solved in a specially defined (see the generalized Neumann BC in Section C of Supplementary Material) closed-system basis taking into account the Hartree potential $\phi^H(\bm{r}_i)$ and the exchange and correlation potential $\phi^{XC}(\bm{r}_i)$. Second, the local density of states (LDOS) of the open system, $\rho(\bm{r}_i,E)$, and the electron density, $n(\bm{r}_i)$, are computed using the CBR method for each iteration. Then the electrostatic potential and the carrier density are used to calculate the residuum $F$ of the Poisson equation
\begin{equation}
\big|\big|\bm{F}[\bm{\phi}^H(\bm{r}_i)]\big|\big|=\big|\big|\bm{A}\bm{\phi}^H(\bm{r}_i) - (\bm{n}(\bm{r}_i)-\bm{N}_D(\bm{r}_i)+\bm{N}_A(\bm{r}_i))\big|\big|,
\end{equation}
where $\bm{A}$ is the matrix derived from the discretization of the Poisson equation and $\bm{N}_D$ and $\bm{N}_A$ are the total donor and acceptor doping densities arrays, respectively. If the residuum is larger than a predetermined threshold $\epsilon$, the Hartree potential is updated using the predictor-corrector method, together with the Anderson mixing scheme. Using the updated Hartree potential and the corresponding carrier density, the exchange-correlation is computed again for the next step, and an iteration of Schrodinger-Poisson is repeated until the convergence is reached with $\big|\big|\bm{F}[\bm{\phi}^H(\bm{r}_i)]\big|\big|<\epsilon=10^{-6}$~eV. Further details of the methodology are included in the Supplementary Material.
In our simulations we have utilized a 3D real-space model, with a discretization size of 0.2~nm along all directions, thus with about $10^{6}$ real-space grid points, and around 3,000 energy points were used. The CBR algorithm automatically ascertains that out of more than 1,000,000 eigenstates only about 700 ($<0.1\%$) of lowest-energy states is needed, which is generally determined by the material properties (e.g. doping level), but not the device size. We have also employed the standard values of the inertial effective mass tensor for electrons, $m_l = 0.98 \times m_e$, $m_t = 0.19 \times m_e$, the dielectric constant of Silicon, $\epsilon_{Si}=11.7$, and the cryogenic temperature of T$=4.0$~K in all our simulations.

\section{Results and discussion}

\subsection{Conductivity of ideal tunnel junctions}\label{sec:conductivity of ideal tunnel junctions}

\begin{figure}
  \centering
  \includegraphics[width=\linewidth]{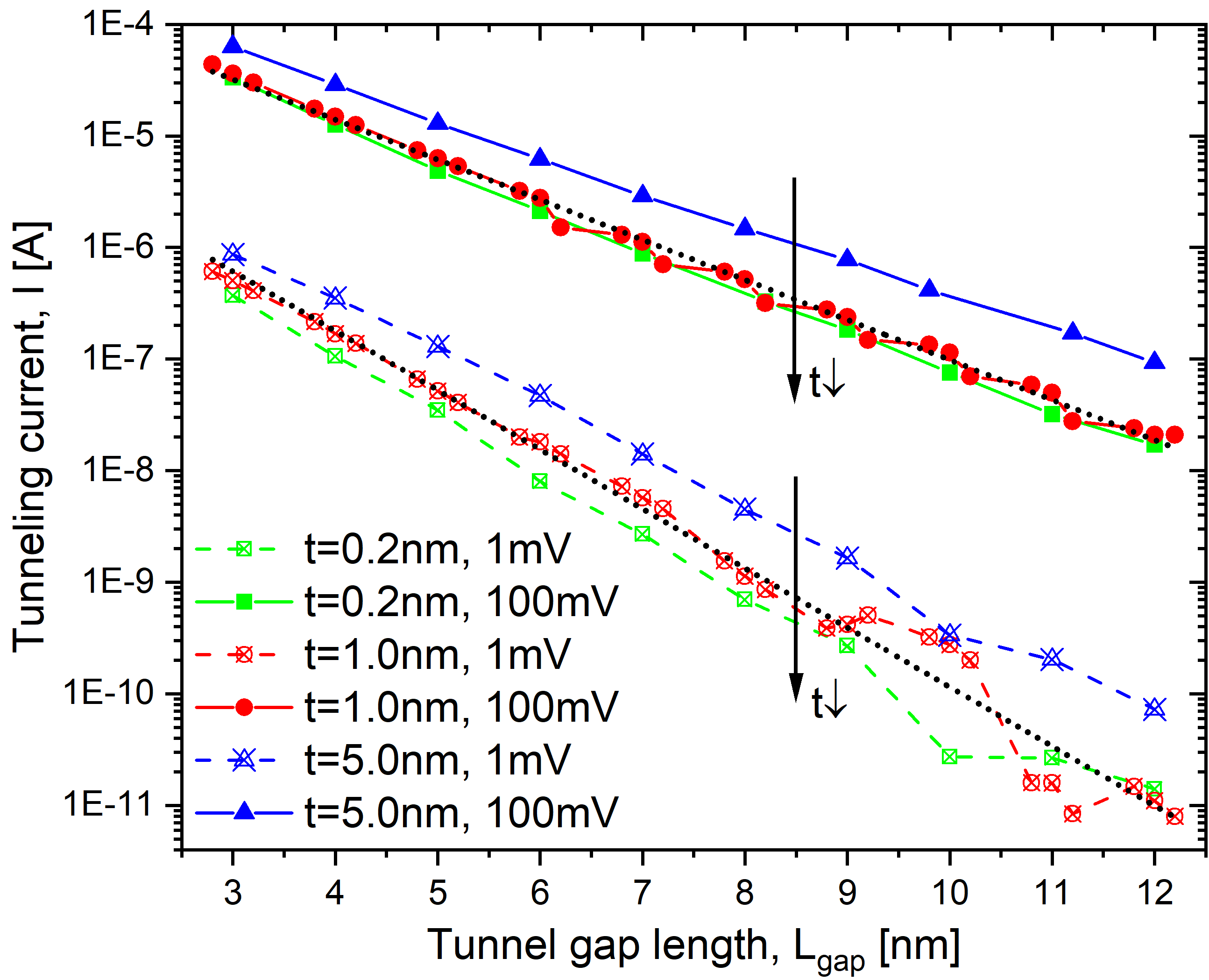}
  \caption{
  \textbf{Characteristic tunneling current curves.} Total current $I$ (semi-logarithmic scale) vs. tunnel gap length $L_{gap}$ for different values of $\delta$-layer thickness $t$ and applied voltages ($1$~mV and $100$~mV). Black dotted lines represent least-square fits to the exponential trend. $N_D=1.0 \times 10^{14}$~cm$^{-2}$ and $N_A=5.0 \times 10^{17}$~cm$^{-3}$.
  }
  \label{fig:I_vs_W_ideal_TJ}
\end{figure}

\begin{figure}
  \centering
  \includegraphics[width=\linewidth]{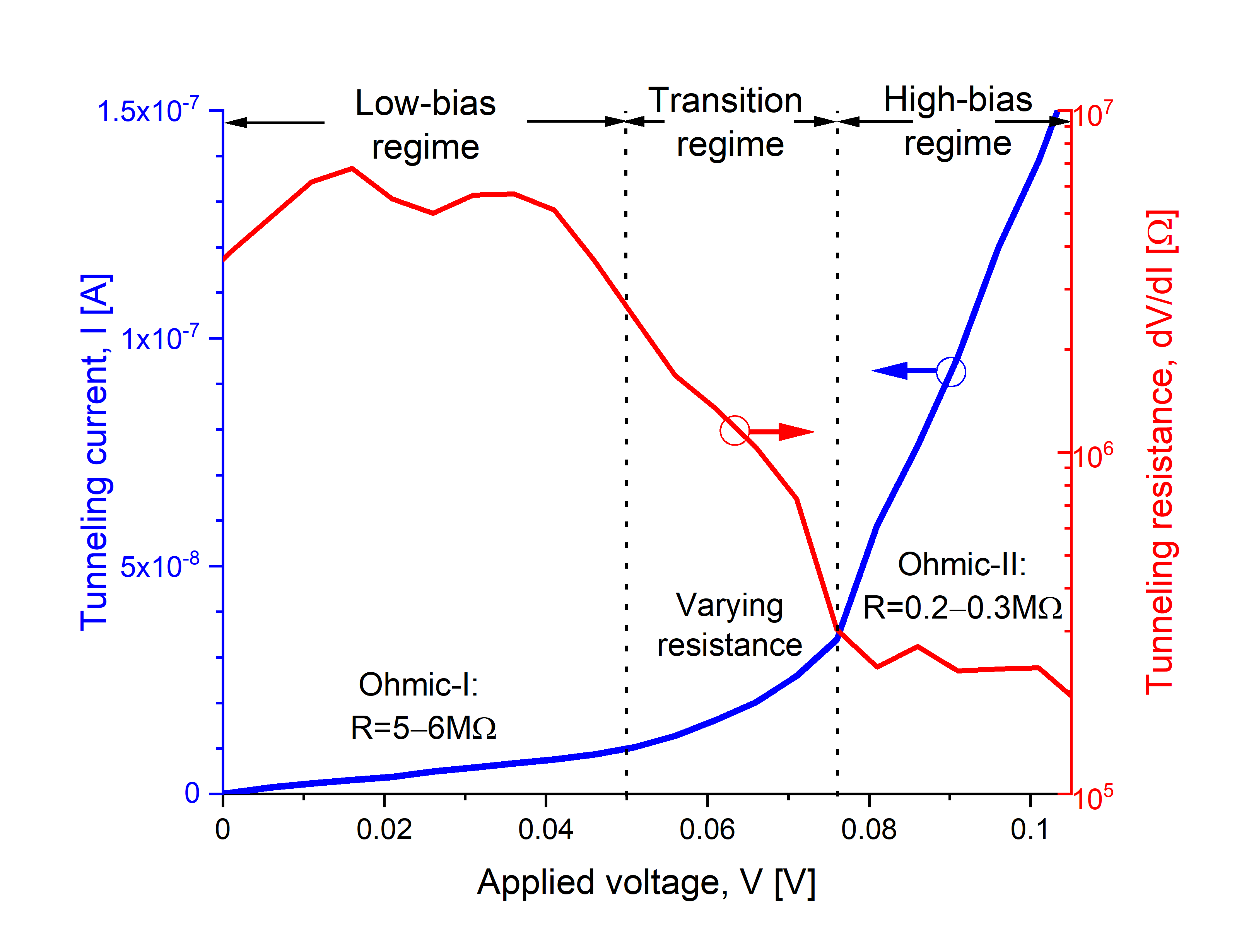}
  \caption{
  \textbf{Two conductivity regimes in $\delta$-layer tunnel junctions.} Total current vs voltage (blue curve, linear scale) and the corresponding differential resistance $dV/dI$ (red curve, semi-logarithmic scale) are shown for $L_{gap}=10$~nm, $N_D=1.0 \times 10^{14}$~cm$^{-2}$, $N_A=5.0 \times 10^{17}$~cm$^{-3}$ and $t=1$~nm.}
  \label{fig:I-V}
\end{figure}

\begin{figure*}
  \centering
  \includegraphics[width=\linewidth]{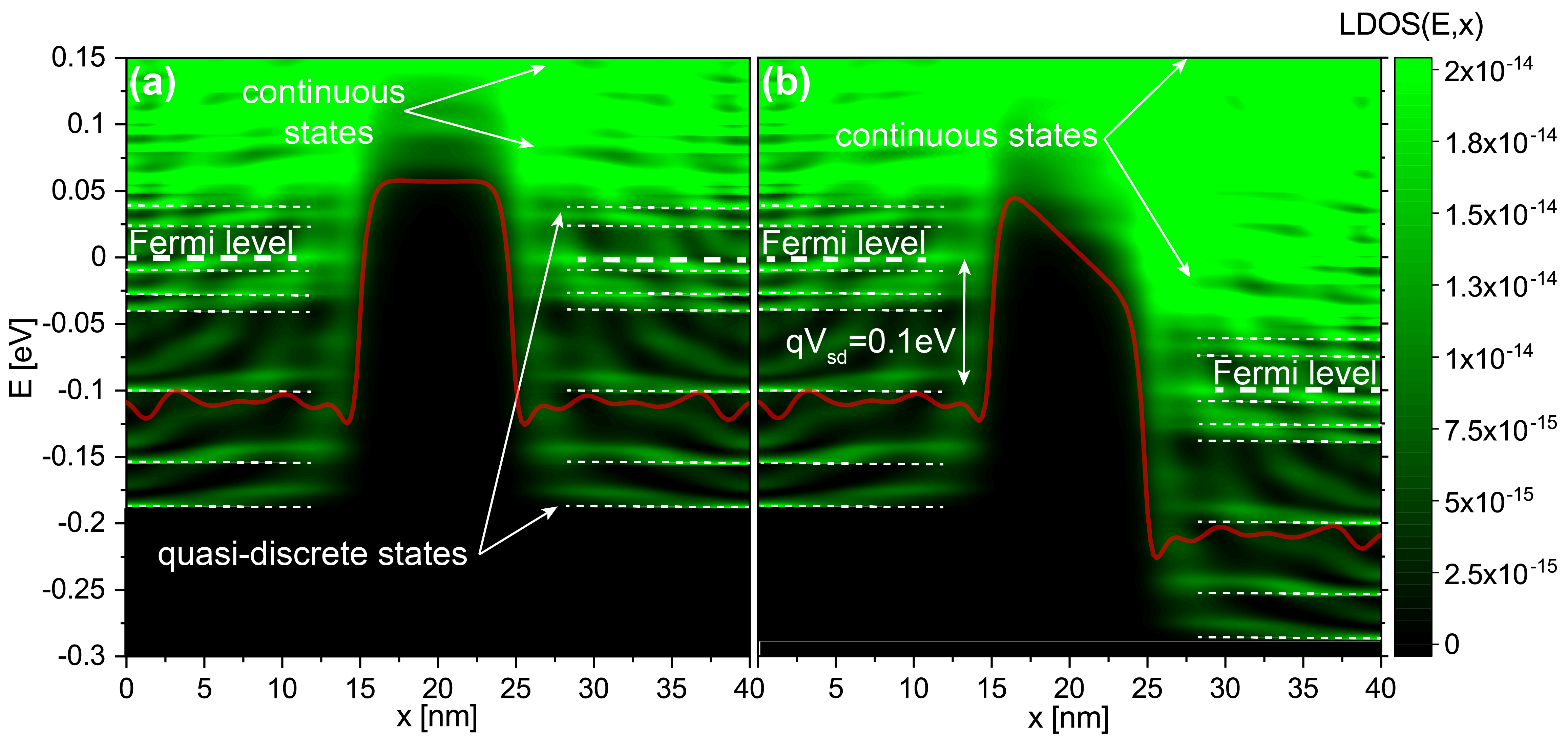} 
  \caption{\textbf{Local Density of States for $\delta$-layer tunnel junctions.} The $LDOS(E,x)$ for a tunnel junction of $L_{gap}=10$~nm is shown in \textbf{a} and \textbf{b} when a voltage of 1~mV and 100~mV is applied to the drain contact, respectively. The Fermi levels indicated in the figures correspond to the Fermi levels of the source and drain contacts. In \textbf{a} and \textbf{b}, the corresponding effective 1D potentials are also shown, calculated by integrating over the (y,z)-plane the actual charge self-consistent 3D potentials weighted with the electron density. $N_D=1.0 \times 10^{14}$~cm$^{-2}$, $N_A=5.0 \times 10^{17}$~cm$^{-3}$, and $t=1$~nm.}
  \label{fig:LDOS_x}
\end{figure*}

In our previous work for infinite-width ($W \to \infty$) $\delta$-layer systems \cite{Mamaluy:2021}, we demonstrated that the distribution of dopants along the confinement z-axis for a fixed sheet doping density ($N_D^{(2D)}$) significantly affects the current. Conductivity decreases for sharper $\delta$-layer doping profiles, which create strong transverse electric fields in their vicinity, while it increases for broader $\delta$-layers doping profiles, which conversely create weaker transverse electric fields. We also report in this work that the same trend is observed for finite-width $\delta$-layer tunnel junctions. In Fig.~\ref{fig:I_vs_W_ideal_TJ}, the tunneling current $I$ vs the tunnel gap length $L_{gap}$ for an ideal $\delta$-layer tunnel junction (see \ref{fig:TJ_model}~\textbf{a}) is included for different $\delta$-layer thicknesses and two voltages: 1mV and 100 mV. As shown, the tunneling current decreases as the $\delta$-layer thickness decreases for a fixed sheet doping density (i.e. the total charge density is kept constant). We also find that the overall $I$~vs~$L_{gap}$ trend is practically exponential for all tunnel gap lengths $L_{gap}=0,...,12$~nm, i.e. $\ln{I}\sim\ln{I_{L_{gap}=0}}-L_{gap}/B_{voltage}$, where $I_{L_{gap}=0}$ is the current when $L_{gap}=0$, $L_{gap}$ is the tunnel gap length and $B_{voltage}$ is a proportional constant related with the barrier height. As a guide to the eyes, the exponential $I$~vs~$L_{gap}$ trend is included in the figure as black dotted lines for $t=1$~nm. However, a deviation from the exponential trend can be noticed for large tunnel gaps $L_{gap}>7$~nm and an applied voltage of $1$~mV (dashed lines in the figure). Conversely, for a voltage of $100$~mV (continous lines in the figure), the deviation from the ideal trend vanishes and the overall trend is exponential, thus shedding light on two conductivity regimes for $\delta$-layer tunnel junctions\cite{Mendez_CS:2022}: low- and high-voltage regimes. As reported in \cite{Mendez_CS:2022}, the deviation from the ideal trend is the result of the quantized conduction band in both $\delta$-layers (left and right) and a certain mismatch between the left and right states, which importantly can only occur for low applied voltages. 

Fig.~\ref{fig:I-V} shows the I-V characteristic curve for an ideal tunnel junction of length $L_{gap}=10$~nm and $t=$1~nm. We can discern two ohmic behaviors, elucidating again the existence of two distinct conductivity regimes corresponding to the low and high voltage: the first one, between $0$~V and $0.04$~V, with a resistance of approximately $5-6$~M$\Omega$; the second one, above $0.08$~V, with a resistance of $0.2-0.3$~M$\Omega$. The resistance in the low-voltage regime is one order of magnitude higher than in the high-voltage regime. Between these two regimes, approximately between $0.04$~V and $0.08$~V, there is a transition region over which the resistance is reduced. The
tunneling resistance in the low-bias regime agrees very well with the measured resistances for tunnel junctions in \cite{Donnelly:2023} for the same regime \cite{Supplementary_material}. Additionally, we note that the existence of two conductivity regimes in $\delta$-layer tunnel junctions agrees well with recent experimental I-V measurements \cite{Donnelly:2021, Donnelly:2023}.

To get a better understanding of the two conductivity regimes (see Fig.~\ref{fig:I-V}) and the strong influence of the quantized conduction band on the tunneling current for low voltages (see the oscillations for $L_{gap}>7$~nm in Fig.~\ref{fig:I_vs_W_ideal_TJ}), we examine the local density of states (LDOS), which represent the available states that can be occupied by the free electrons in space-energy dimension. For very low temperatures the states below Fermi level are occupied, whereas the states above Fermi level are unoccupied. Fig.~\ref{fig:LDOS_x} shows the LDOS along x-direction for a tunnel junction of $L_{gap}=10$~nm when a voltage of 1~mV in \textbf{a} and 100~mV in \textbf{b} is applied to the drain contact. Additionally, the corresponding effective 1D potential is included in the figure, exhibiting a tunnel barrier height of approximately 55~meV for the equilibrium case, which is in excellent agreement with the estimation height obtained from the measured I-V curve using the WKB approximation for direct tunneling resistance \cite{Donnelly:2021} and the tight-binding calculations for the barrier height in \cite{Donnelly:2023}. Firstly, as shown in Fig.~\ref{fig:LDOS_x}, the low-energy LDOS are strongly quantized, highlighted with dashed lines in the figure. This strong quantization (or similarly the presence of quasi-discrete states) in the low energies, is the result of the strong confinement of the electrons in the z-direction due to the ultra-thin $\delta$-layer. The presence of discrete states have been observed experimentally in several high resolution ARPES measurements for $\delta$-layers in silicon \cite{Holt:2019,Mazzola:2019}. On the contrary, for high energies, the LDOS are practically continuous in space-energy, thus these states are not quantized. When a voltage is applied to the drain contact, the Fermi level corresponding to the drain contact is reduced, resulting in lowering the energies of all states in the right side as well. As a result, new unoccupied states in the right $\delta$-layer will be available to be occupied by the tunneling electrons coming from the left $\delta$-layer. When a low drain voltage is applied, $<45-50~\text{mV}$, only the unoccupied quantized states in the right $\delta$-layer will play a role in the tunneling process. If the occupied quasi-discrete states near the Fermi level in the left side align with the unoccupied quasi-discrete states in the right side, it will result in a considerable increase of the tunneling current as shown in Fig.~\ref{fig:I_vs_W_ideal_TJ} for $L_{gap}=10$~nm and $t=1$~nm. Conversely, if the overlap is minimum, as happen for $L_{gap}=11$~nm and $t=1$~nm, the tunneling current will be reduced. For low biases, this alternating mismatch can only exist for sufficiently large tunnel gaps, $L_{gap}>7$~nm, because the coupling of the left and right $\delta$-layer wave-functions for narrow tunnel gaps ($L_{gap}<7$~nm) equalizes the electron states on both sides, increasing the overlap and thus eliminating the mismatch. When a high bias is applied, as in Fig.~\ref{fig:LDOS_x}~\textbf{b}, it makes the continuous unoccupied high-energy states in the right side available for tunneling from the left side, thus diminishing the influence of the conduction band quantization on the current, as can be seen in $I$ vs $L_{gap}$ plots in Fig.~\ref{fig:I_vs_W_ideal_TJ} for $100$~mV.

In the following we will evaluate the effect of diverse imperfections in $\delta$-layer tunnel junctions on the tunneling current. We therefore remark the importance of the evaluation of the tunneling rate for both conductivity regimes: low- and high voltage regimes. For low voltages, strong quantization effects on the tunneling current are expected, especially for large tunnel gaps, because of the quantized low-energy conduction band; thus, it will be reflected in non-monotonic or oscillated characteristic I-V curves. Conversely, in the high-bias regime, no significant influence of the low-energy conduction band quantization is expected on the tunneling current.

\subsection{Effects of $\delta$-layer thickness deviations from mono-atomic layer}\label{sec:Variation of the delta-layer thickness}

\begin{figure}
  \centering
  \includegraphics[width=\linewidth]{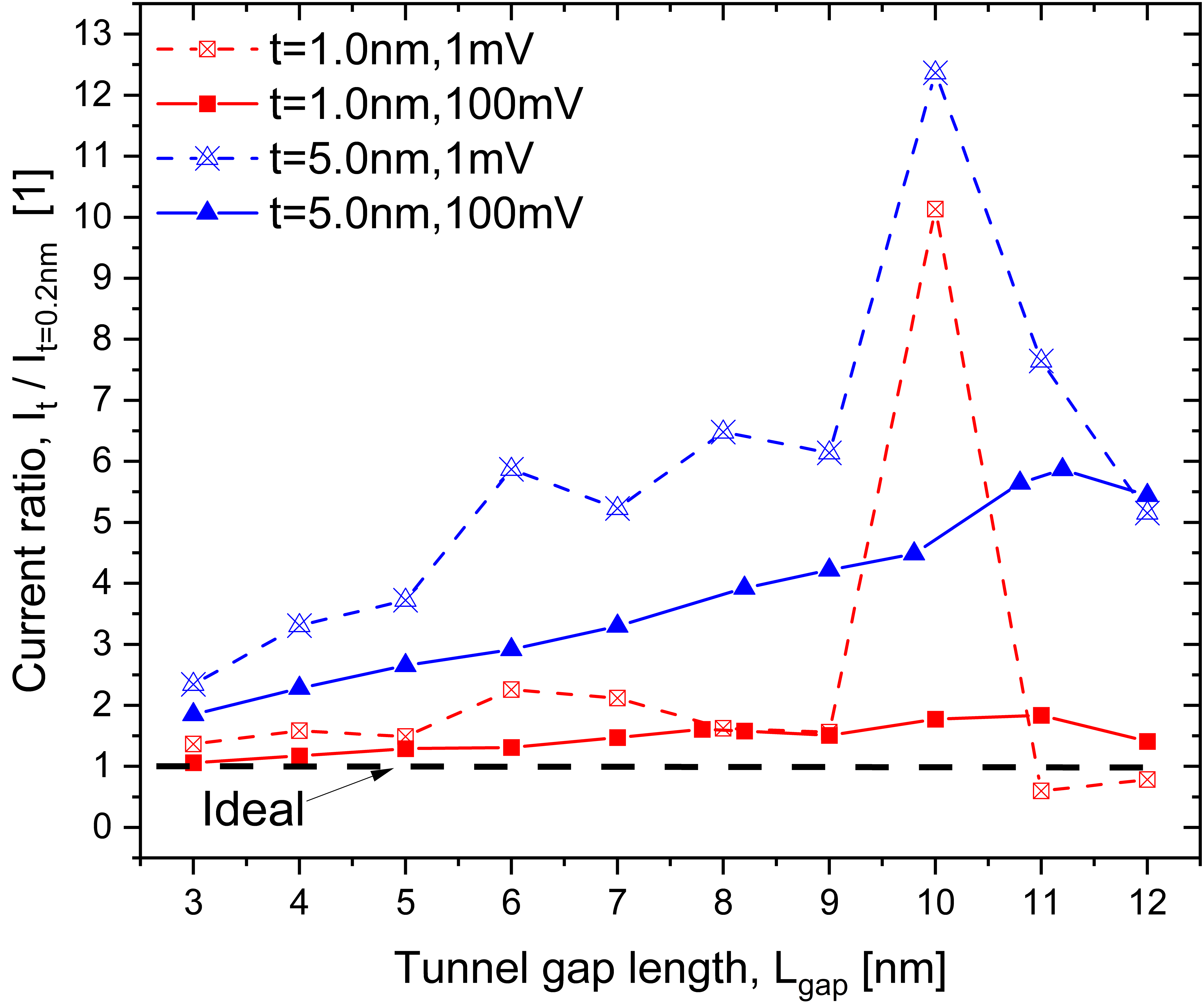} 
  \caption{
  \textbf{Effect of $\delta$-layer thickness variation.} Current ratio, $I_{t}/I_{t=0.2~\text{nm}}$, vs. tunnel gap length for different deviations of the $\delta$-layer thickness from the "ideal" mono-atomic layer.
  }
  \label{fig:current ratio thickness variation}
\end{figure}

We first investigate the effects of $\delta$-layer thickness variation on the tunneling rate. Fig.~\ref{fig:current ratio thickness variation} shows the current ratio between two different $\delta$-layer thicknesses ($t=1,~5$~nm) with respect to the ideal mono-atomic layer, which is approximately 0.2~nm, in dashed lines for a low-bias of $1$~mV and in continuous lines for a high-bias of $100$~mV. Our results suggest that the tunneling rate increases approximately up to two times for a broadening of the $\delta$-layer of $1$~nm. Interestingly, the effect is roughly constant for the whole considered tunnel gap $L_{gap}$ range. However, for even wider $\delta$-layers, the tunneling rate further increases, between two and seven times for a effective thickness of $5$~nm, developing now a stronger dependence on the tunnel gap length. Furthermore, for large tunnel gaps in the low-bias regime ($1$~mV), the strong non-monotonic dependence of the current ratio on the gap length is the result of the space-energy quantization of the conduction band and the mismatch in the overlapping between the states in the left and right $\delta$-layers, discussed in Sect.~\ref{sec:conductivity of ideal tunnel junctions}. More specifically, the peak at $L_{gap}=10$~nm is the result of maximum overlap between occupied quasi-discrete states from the left $\delta$-layer with unoccupied quasi-discrete states from the right $\delta$-layer for $\delta$-layers thicker than monoatomic layers. It is worth noting that the energy levels of these quasi-discrete states exhibit a dependence on multiple factors, including the doping density, width and thickness, and length of the tunnel gap. When a higher voltage is applied (e.g. for $100$~mV), it results in a lower current ratio, and the quantization effects diminish as we discussed in Sect~\ref{sec:conductivity of ideal tunnel junctions}.

\subsection{Effects of  tunnel gap length variations and interface roughness}\label{sec:variation of the tunnel gap length}

\begin{figure}
  \centering
  \includegraphics[width=\linewidth]{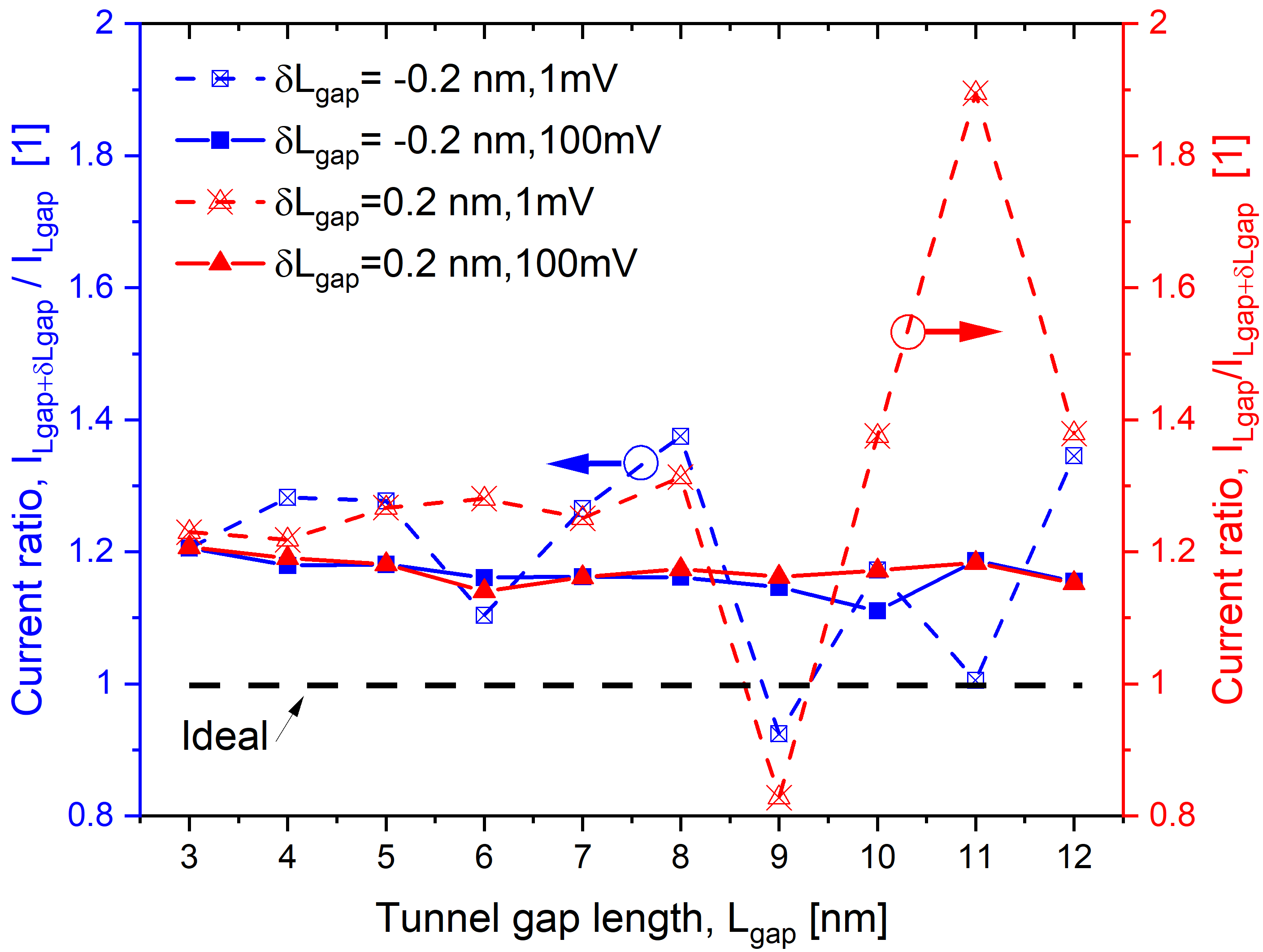} 
  \caption{
  \textbf{Effect of tunnel gap length variation.} Current ratio, $I_{L_{gap} + \delta L_{gap}}/I_{L_{gap}}$ vs. tunnel gap length, $L_{gap}$, for distinct applied voltages. $t=1.0$~nm, $N_D=1.0 \times 10^{14}$~cm$^{-2}$ and $N_A=5.0 \times 10^{17}$~cm$^{-3}$.
  }
  \label{fig:current_ratio_geometry_variation}
\end{figure}

\begin{figure}
  \centering
  \includegraphics[width=\linewidth]{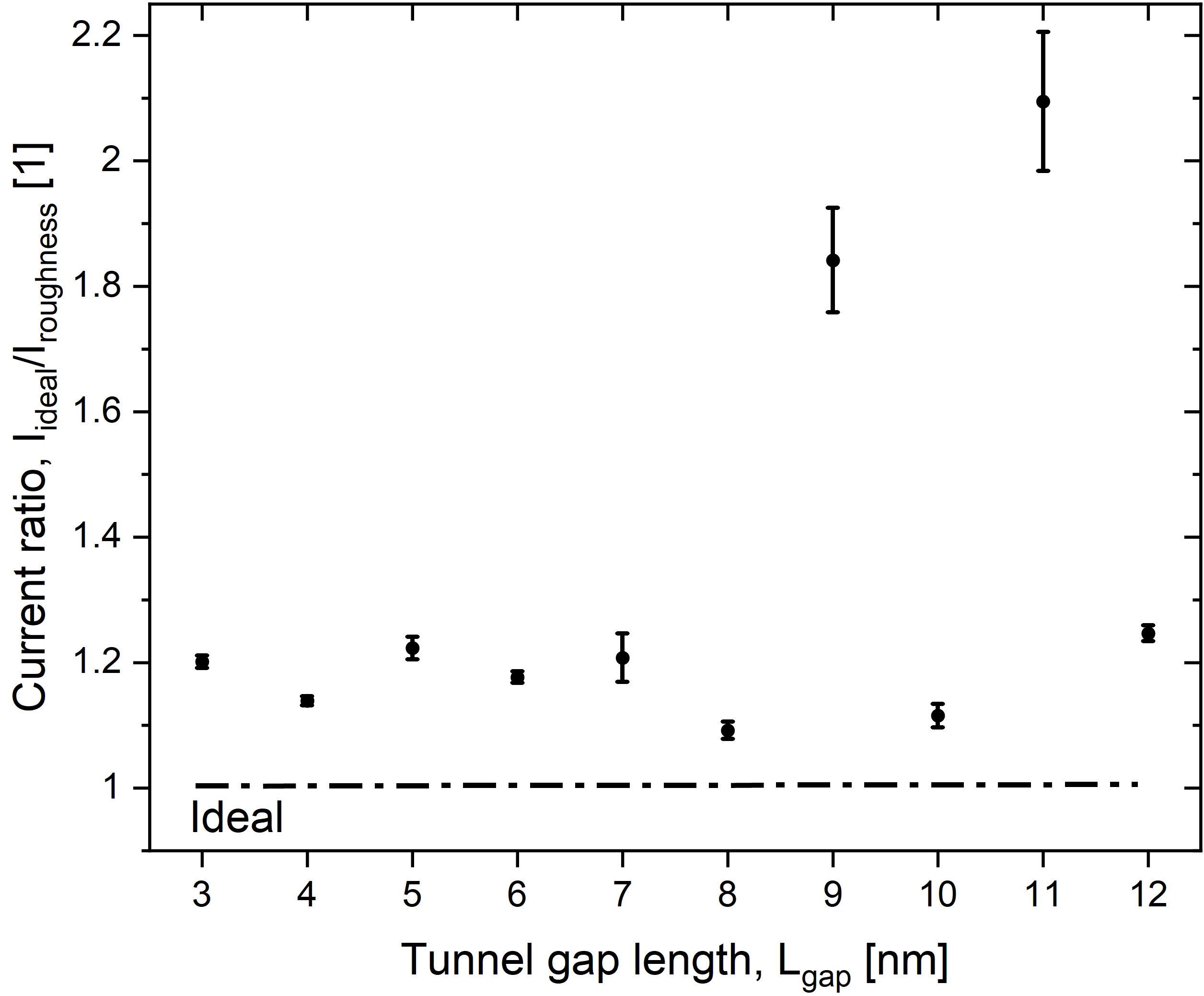}
  \caption{
  \textbf{Effect of interface roughness.} Current ratio, $I_{ideal}/I_{roughness}$, vs. tunnel gap length, $L_{gap}$, for an applied bias of $1.0$~mV. For each tunnel gap length, different roughness sizes have been considered, ranging $d_1=0.8-2.0$~nm and $d_2=0.6-3.4$~nm (see Fig.~\ref{fig:TJ_model}~\textbf{b}). The dots represent the average and the bars represent the dispersion of the values. $t=1.0$~nm, $N_D=1.0 \times 10^{14}$~cm$^{-2}$ and $N_A=5.0 \times 10^{17}$~cm$^{-3}$.
  }
  \label{fig:current_ratio_roughness}
\end{figure}

Next we will assess how the variation of the tunnel gap length and the roughness of the $\delta$-layer might affect the tunneling rate. As a first approximation, the interface roughness can be modeled as a uniformed increase or reduction of the average gap length $\langle L_{gap} \rangle$. We evaluate  in Fig.~\ref{fig:current_ratio_geometry_variation} the change in the tunneling current for small uniform variations, such as $L_{gap} + \delta L_{gap}$ with $\delta L_{gap} = \pm 0.2$~nm. This small perturbation is of the order of the stochasticity of APAM chemistry. Our simulations suggest that a small variations of the tunnel gap length (of the order of $\pm 0.2$~nm) can lead to a current change of around 20$\%$ with respect to the ideal (or designed) length for the whole studied range. A reduction of the tunnel gap length evidently results in an increase of the tunneling rate, while an increase of the effective gap length leads to a decrease of the tunneling rate by a similar magnitude. 
For large tunnel gaps in the low-bias regime (see dashed lines for $1$~mV), our results similarly present the expected non-monotonic behaviour due to the quantization effect, but this effect vanishes when higher voltage is applied (see continuous lines for $100$~mV).

A second-order analysis of the edge roughness can be performed assuming that the average $\langle L_{gap} \rangle$ value does not change due to the roughness. In this case, we model the roughness as periodic protrusion of size $d_1 \times d_2 \times t$, with a periodicity of $2 \times d_2$, as shown in Fig.~\ref{fig:TJ_model} \textbf{b}, instead of the uniformed variation evaluated above. To maintain $\langle L_{gap} \rangle$ constant we have only considered in-phase roughness, i.e. the protrusion of the left $\delta$-layer is exactly a mirror of the right side. The analysis of out-of-phase roughness is not the scope of this work and it will be further investigated outside. In our analysis, we have considered different roughness sizes, varying the parameter $d_1$ from $0.8$~nm to $2.0$~nm, and the parameter $d_2$ from $0.6$~nm to $3.4$~nm. Fig.~\ref{fig:current_ratio_roughness} includes, for all evaluated roughness sizes, the current ratio between the "non-ideal" tunnel junction device (with edge roughness) and the "ideal" device (without any roughness) for a voltage of $1$~mV. The simulations indicate that edge roughness reduces the tunneling rate between $6\%$ and $20\%$ for almost the whole considered gap range, predicting very similar magnitude of the tunneling rate change as the uniformed variation of the tunnel gap length. One can also notice that for large tunnel gaps, especially for $L_{gap}=9-10$~nm, the tunneling rate is even further reduced up to 2.2 times due to the mentioned quantization of the conduction band.

\subsection{Effects of charged impurities in the tunnel gap}\label{sec:A charged impurity in the tunnel gap}

\begin{figure*}
\includegraphics[width=\linewidth]{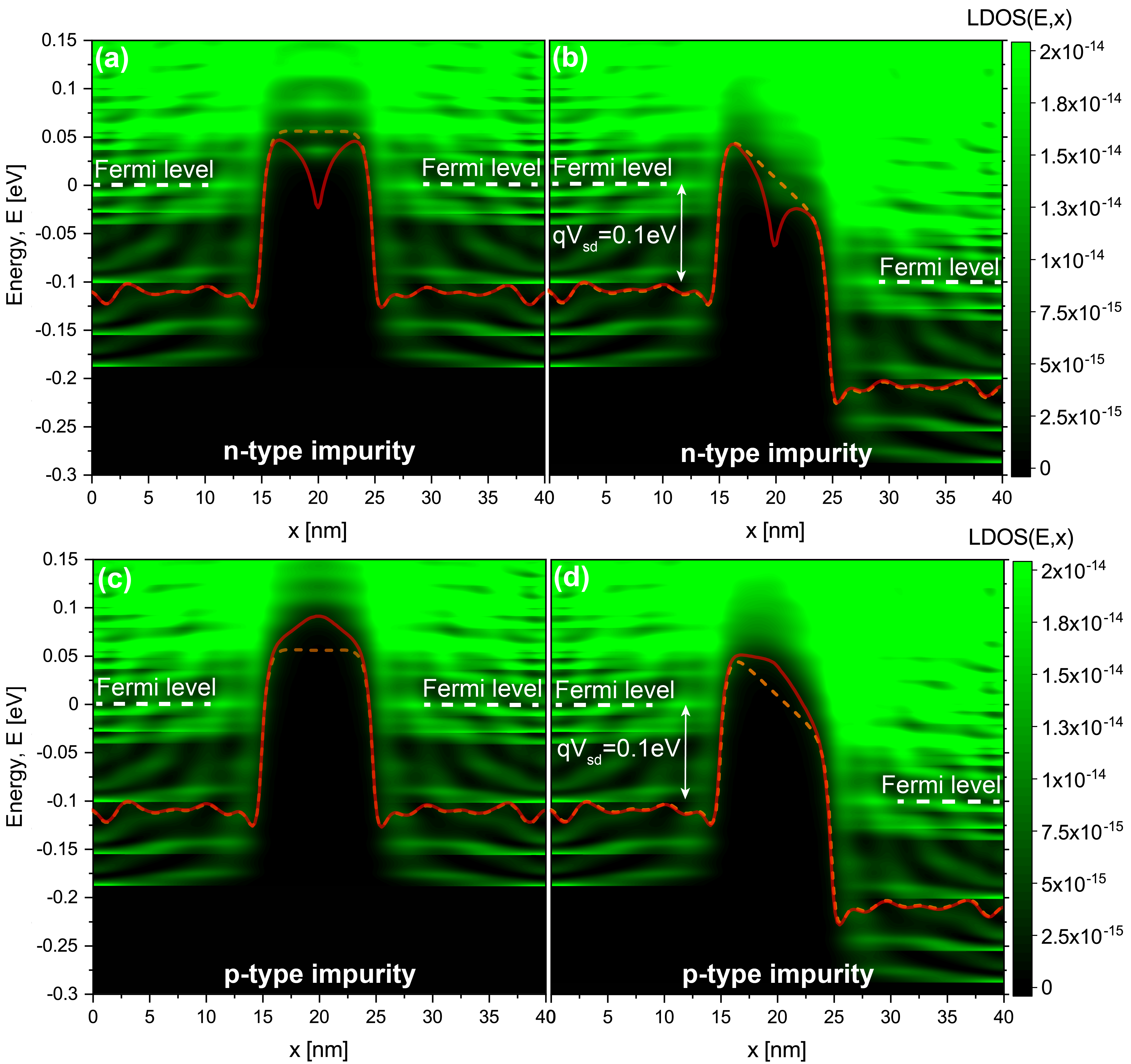}
\caption{\textbf{Local Density of States for $\delta$-layer tunnel junctions with impurities.} The $LDOS(E,x)$ for a tunnel junction of $L_{gap}=10$~nm is shown: for a n-type impurity when a voltage of 1~mV and 100~mV is applied to the drain contact in \textbf{a} and \textbf{b}, respectively; for a p-type impurity when a voltage of 1~mV and 100~mV is applied in \textbf{c} and \textbf{d}, respectively. In all figures, the corresponding effective 1D potentials are also shown in red color, calculated by integrating over the (y,z)-plane the actual charge self-consistent 3D potentials weighted with the electron density. The effective 1D potential for the ideal tunnel junction is also included in orange dashed lines for comparison purpose. $L_{gap}=10$~nm, $t=1.0$~nm, $N_D=1.0 \times 10^{14}$~cm$^{-2}$, and $N_A=5.0 \times 10^{17}$~cm$^{-3}$.}
\label{fig:LDOS_impurities}
\end{figure*}

\begin{figure*}
\centering
\includegraphics[width=\linewidth]{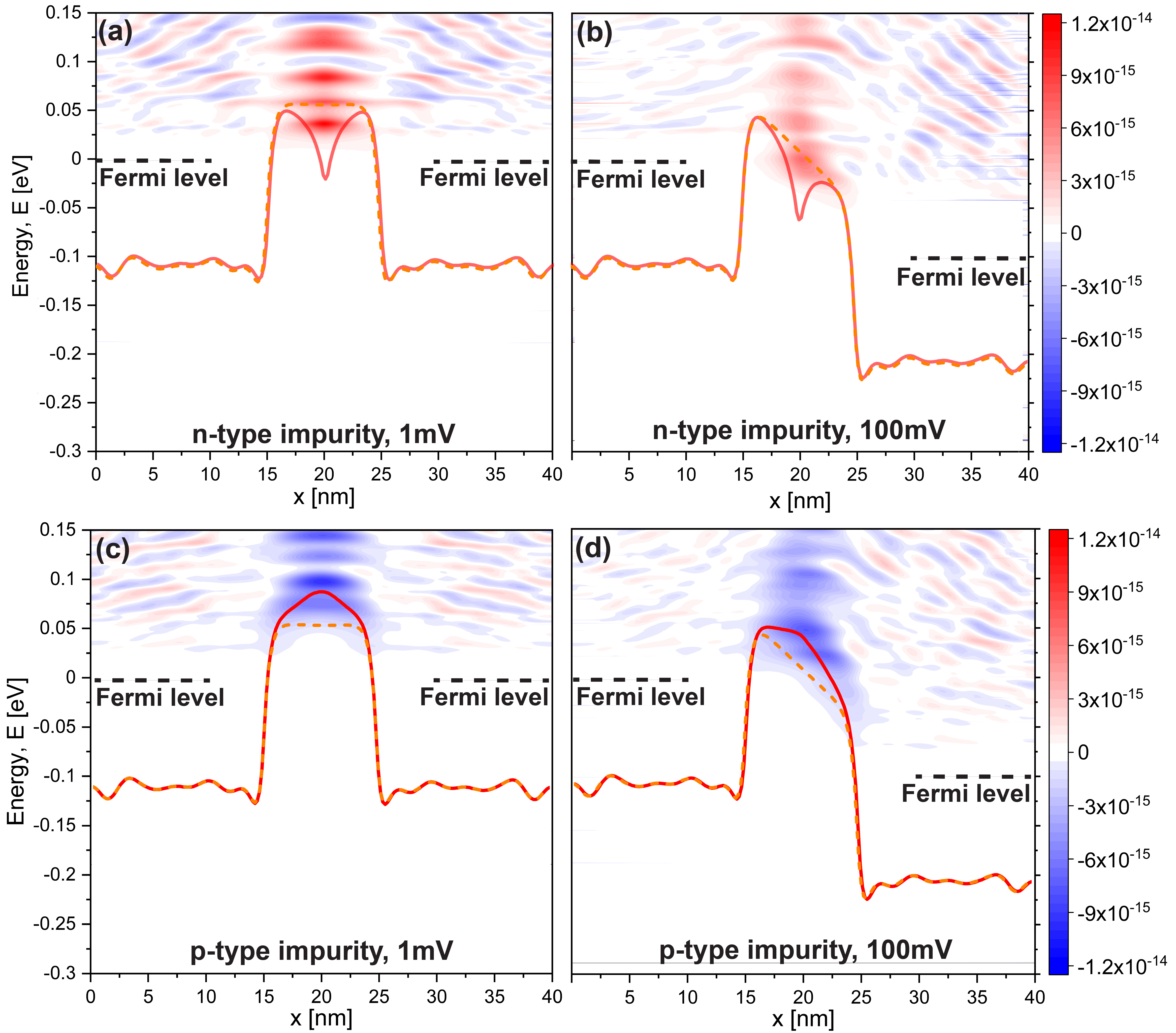} 
\caption{ \textbf{Localized states created or depleted by a charged impurity.} Shown is the $LDOS(E,x)$ difference for a tunnel junction of $L_{gap}=10$~nm between: n-type impurity in the middle of the tunnel gap and without impurity in \textbf{a} and \textbf{b} for an applied voltage of 1~mV and 100~mV, respectively; p-type impurity and without impurity for an applied voltage of 1mV and 100~mV in \textbf{c} and \textbf{d}, respectively. The corresponding effective 1D potentials are also shown in red color, together with the effective 1D potential for the ideal case in orange color. $N_D=1.0 \times 10^{14}$~cm$^{-2}$, $N_A=5.0 \times 10^{17}$~cm$^{-3}$, and $t=1$~nm.}
\label{fig:LDOS_impurities_difference}
\end{figure*}

\begin{figure}
 \includegraphics[width=\linewidth]{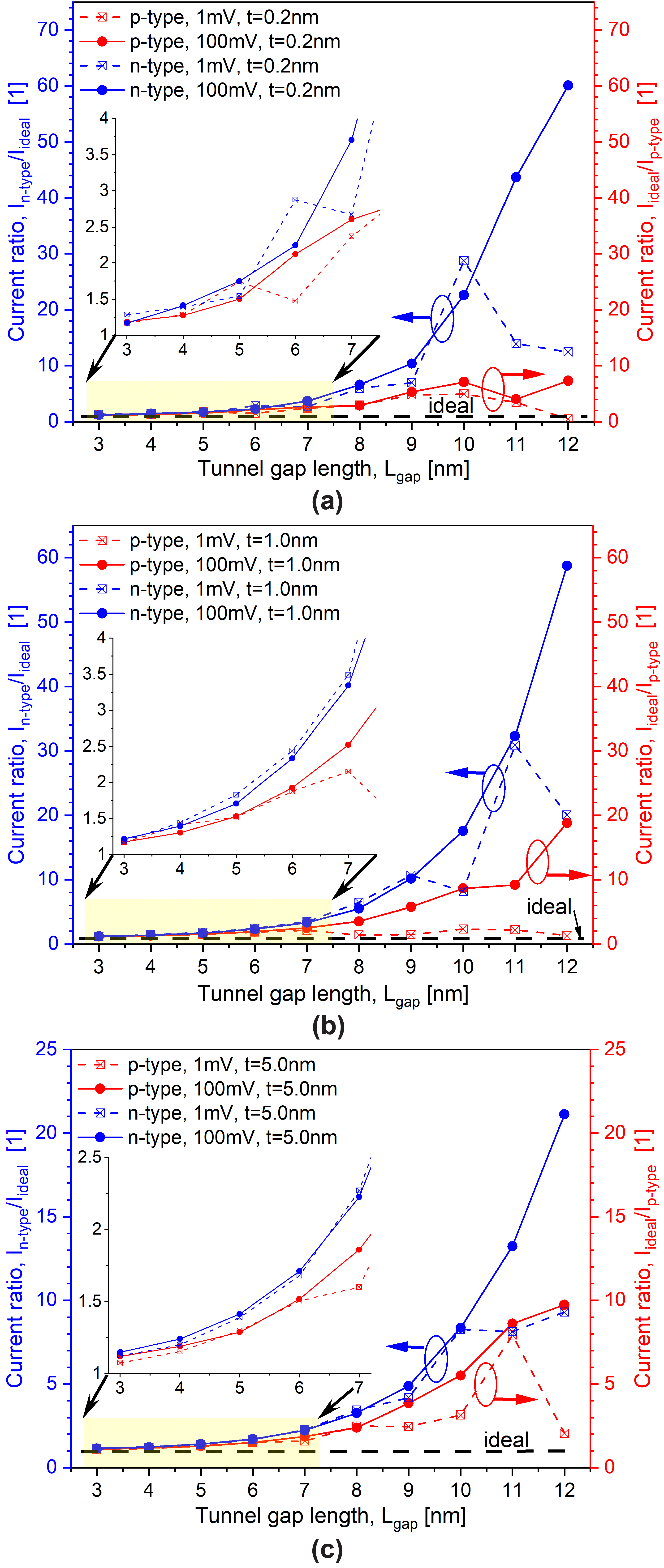}
 \caption{\textbf{Effect of charged impurities.} Current ratio, $I_{non-ideal}/I_{ideal}$, vs. tunnel gap length, $L_{gap}$, for tunnel junctions with a single n-type and p-type impurities in the intrinsic gap: \textbf{a} t=0.2~nm; \textbf{a} t=1.0~nm; and \textbf{c} t=5.0~nm. The insets in \textbf{a}, \textbf{b} and \textbf{c} are a zoom of the result within the range between 3~nm and 8~nm. $N_D=1.0 \times 10^{14}$~cm$^{-2}$ and $N_A=5.0 \times 10^{17}$~cm$^{-3}$.}
\label{fig:current_ratio_impurities}
\end{figure}

In the following, we evaluate the effects on the tunneling rate by the presence of a single charged impurity in the tunnel gap, assuming a point-charge distribution for all charged impurities regardless of the specific atomic species, as shown in Fig.~\ref{fig:TJ_model} \textbf{c}. For the simulation set-up, we place in the center of the tunnel gap either a n-type impurity (e.g. a phosphorus atom) or a p-type impurity (e.g. an aluminium atom). The impurities are modeled by approximating a point charge with a density of (positive or negative) $4.6 \times 10^{21}$~cm$^{-3}$ homogeneously distributed in a total volume of (0.6~nm)$^3$. While in this work we restrict our analysis to the center-gap location, the influence of other locations may be also interesting to investigate since the free electrons in $\delta$-layer systems form distinct conducting layers perpendicular to the confinement direction, thus signaling a highly non-homogeneous electron density distribution \cite{Mamaluy:2021,Mendez:2020}.

Fig.~\ref{fig:LDOS_impurities} shows the LDOS of a $\delta$-layer tunnel junction with a single n-type impurity for an applied voltage of 1~mV in \textbf{a}  and for 100~mV in \textbf{b}, and with a single p-type impurity for 1~mV in \textbf{c} and for 100~mV in \textbf{d}. In addition, Fig.~\ref{fig:LDOS_impurities_difference} shows the LDOS difference for both applied voltages between the ideal tunnel junction (i.e. the LDOS shown in Fig.~\ref{fig:LDOS_x}) and the junction with the impurity (i.e. the LDOS shown in Fig.~\ref{fig:LDOS_impurities}), therefore it represents the localized states created (in red color) or depleted (in blue color) by the impurity. As Fig.~\ref{fig:LDOS_impurities}~\textbf{a} and \textbf{b}  reveal, an n-type impurity in the middle of the tunnel gap creates unoccupied states above the Fermi level within the tunnel gap, i.e. between $x=15-25$~nm, as one might discern comparing panels \textbf{a} and \textbf{b} in Fig.~\ref{fig:LDOS_x} and Fig.~\ref{fig:LDOS_impurities}. The new available states are clearly shown in red color in panels \textbf{a} and \textbf{b} in Fig.~\ref{fig:LDOS_impurities_difference}. On the contrary, a p-type impurity depletes unoccupied available states above the Fermi level as shown in Fig.~\ref{fig:LDOS_impurities}~\textbf{c} and \textbf{d} within the tunnel gap ($x=15-25$~nm). The depletion of the states due a p-type impurity can be also seen in Fig.~\ref{fig:LDOS_impurities_difference}~\textbf{c} and \textbf{d} in blue color since we are representing the LDOS difference between the ideal tunnel junction and the one with the p-type impurity. Interestingly, the presence of the impurity can be sensed even far from the impurity position as Fig.~\ref{fig:LDOS_impurities_difference} also reveals for both impurity types. The impurity creates a quantized perturbation in the unoccupied available states (above the Fermi level) even far from the impurity location, shown as ripples between $x=0-15$~nm and $x=15-40$~nm above the Fermi level in the figure. However, the intensity of the perturbations vanish as we move away from the impurity location.

In the semi-classical picture, the energy difference between the peak of the effective electrostatic potential and the Fermi level is the energy barrier which the electrons have to overcome to tunnel from one $\delta$-layer to the other. Because of the presence of the impurity in the middle of a nano-scale gap, the effective electrostatic potential is obviously affected. For an n-type impurity in the middle of tunnel gap, the height of the electrostatic barrier is reduced, as shown in Fig.~\ref{fig:LDOS_impurities}~\textbf{a}, in which the height of the electrostatic barrier without impurity (the orange dashed curve) is slightly higher than the one with an n-type impurity (the red continuous curve).  In addition, the n-type impurity creates a dip in the electrostatic potential due to the donor atom. For a p-type impurity, acceptor atom, the depletion of the states above the Fermi level, in turn, increases the height of the energy barrier in the electrostatic potential with respect to the ideal case, as depicted in Fig.~\ref{fig:LDOS_impurities}~\textbf{c}, where the height of the electrostatic potential for the ideal tunnel junction (red continuous curve) is higher than the electrostatic potential for the tunnel junction with the impurity (orange dashed curve).

Fig.~\ref{fig:current_ratio_impurities} shows the current ratio versus the tunnel gap length $L_{gap}$ for both impurity types and voltage regimes ($1$~mV and $100$~mV) in \textbf{a}, \textbf{b} and \textbf{c} for a $\delta$-layer thicknesses of $t=0.2,~1.0,~5.0$~nm, respectively. These simulations first reveal that a n-type impurity increases the tunneling rate, whereas a p-type impurity decreases the tunneling rate. This result can be explained in two different, but related ways. The first one, which corresponds to a semi-classical picture, is by examining the electrostatic potentials shown in Fig.~\ref{fig:LDOS_impurities} and discussed above: an n-type impurity decreases the barrier height (i.e. the energy difference between the peak of the electrostatic potential and the Fermi level), whereas a p-type impurity increases the barrier height. Then the tunneling current increases or decreases according to the change of the barrier height. The other way is to examine the unoccupied and occupied states in the LDOS for the conduction band. When we apply a positive voltage to the drain contact, the right Fermi level and occupied and unoccupied states move down, falling below the Fermi level of the source. As result, electrons injecting from the source can tunnel into these available states. The presence of an n-type impurity creates unoccupied states within the tunnel gap, just above the Fermi level in equilibrium. Similarly, when a voltage is applied, these available states might also fall within the source and drain Fermi energy levels and, therefore, they became as intermediate states in which electrons can tunnel in and out, reducing then the tunneling length (i.e. the width of the barrier in the semiclassical picture) and, therefore, increasing the tunneling rate. On the contrary, the presence of a p-type impurity creates a depletion of the states within the tunnel gap, as shown panels \textbf{c} and \textbf{d} in Fig.~\ref{fig:LDOS_impurities_difference} as a blue cloud around the effective electrostatic potential, increasing then the effective tunneling length and, therefore, reducing the tunneling current.

We can also observe from Fig.~\ref{fig:current_ratio_impurities} two different behavior, corresponding to narrow tunnel gaps $L_{gap}=3-7$~nm and large tunnel gaps $L_{gap}>7$~nm, in which the deviation of the tunneling current from the ideal one behave differently. For narrow tunnel gaps, the magnitude of the effect on the tunneling rate is very similar for both impurity types and not very pronounced (see the inset figures): up to $3.75$ times increase and $2.5$ times reduction of the tunneling rate for a n-type and p-type impurity, respectively, when the tunnel gap length is $7$~nm. Our simulations also exhibit that there is not significant difference on tunneling rate change in both voltage regimes ($1$~mV for the low-voltage regime and $100$~mV for the high-voltage regime) for narrow tunnel gaps: the change of the tunnel rate with respect to the ideal tunneling current is only slightly higher for higher voltages. The deviation in the magnitude of the effect on the tunneling current between both impurity types starts approximately for $L_{gap}>7$~nm: the effect of a n-type impurity becomes much more relevant than for a p-type impurity, especially for the high-voltage regime ($100$~mV in the figure), in which it is a few times higher than in the low-voltage regime ($1$~mV in the figure). Interestingly, our simulations suggest that the tunneling current can be up to $60$ times higher and $20$ times lower for a single n- and p-type impurity, respectively, for a tunnel gap length of $12$~nm. Finally, we note that the charged impurity is better sensed by thinner $\delta$-layers, as our result indicates the current ratio increases as the thickness of the $\delta$-layer decreases.

\section{Conclusion}\label{sec:conclusion}

We have employed an open-system quantum transport analysis to investigate the effect of diverse imperfections on the tunneling rate in Si: P $\delta$-layer tunnel junctions. These imperfections span from geometry variations of the $\delta$-layer thickness and junction gap length, to the presence of charged impurities, either n-type or p-type, in the intrinsic gap. It is shown that while most of the disorders moderately affect the tunneling rate, a single charged impurity in the tunnel gap can alter the tunneling rate by more than an order of magnitude. Contrary to predictions of semiclassical impurity scattering (mobility-based) models, the  electric sign of impurity plays an important  role  in  the  tunneling rate: the rate of current increase due to an n-type impurity in Si: P $\delta$-layer systems is several times higher than the rate of current decrease for a p-type impurity, especially for large tunnel gap lengths. Similarly, we can extrapolate these findings to other systems such as Si: B $\delta$-layer tunnel junctions, in which the influence of a p-type impurity in the intrinsic gap, instead of an n-type, would result in a dramatic increase of the tunneling current. 

These results immediately suggest that the overall geometric fidelity of the APAM device fabrication can be less important than mitigation of charged impurities nearby the junction, which can lead to a strong change of tunneling rates. APAM-based qubits in particular require having tunneling rates that are tightly controlled between multiple closely-spaced objects. This includes exchange coupling between a pair of donor-based qubits, initialization from a tunnel-coupled single electron transistor, and spin-to-charge conversion from the resonant tunneling of the single electron transistor to the leads in Figure \ref{fig:example of APAM devices}. Importantly, a change in tunneling rates from the fabricated geometry due to impurities can be hard to compensate for using electrostatic gates.  The size of the gates (tens of nm) and the required spacing for them to not leak to one another (tens of nm) is much larger than the length scale for tunnel coupling (a few nm). In the APAM geometry, these gates are few in number - enough to control electrostatics, but too few to control tunneling rates. This leads to the conclusion that adopting APAM processing practices that minimize charged defects is more important than those that preserve the absolute device geometry.

Finally, the extreme sensitivity of $\delta$-layers tunnel junctions on the tunneling current to the presence of charges in the vicinity of the tunnel gap opens a great opportunity to use them for quantum FET-based sensors for biological, chemical and radiation applications. The signal detection (either due to radiation or specific molecules) at the sensing area would be strongly enhanced due to the conduction band quantization created by the highly-confined $\delta$-layers. The sensing area can be placed above the tunnel gap, replacing the traditional gate in a conventional geometry. Contrary to traditional FET-based sensors, instead of needing to accumulate enough charges at the sensing area  to invert the full channel and detect the signal \cite{PNAS-FET-sensor}, quantum FETs based on $\delta$-layers would allow to detect even small signals that correspond to a single charge, thus significantly enhancing the sensitivity with respect to traditional FET-based sensors.

\section*{acknowledgement}\label{sec:acknowledgement}
The authors are thankful to FAIR DEAL's team at Sandia National Laboratories for the discussions during the project's meetings. This work is funded under Laboratory Directed Research and Development Grand Challenge (LDRD GC) program, Project No. 213017, and under Laboratory Directed Research and development (LDRD) program, Project No. 227155, at Sandia National Laboratories. Sandia National Laboratories is a multimission laboratory managed and operated by National Technology and Engineering Solutions of Sandia, LLC., a wholly owned subsidiary of Honeywell International, Inc., for the U.S. Department of Energy’s National Nuclear Security Administration under contract DE-NA-0003525. This paper describes objective technical results and analysis. Any subjective views or opinions that might be expressed in the paper do not necessarily represent the views of the U.S. Department of Energy or the United States Government.

\section*{Author contributions}
J.P. and D.M. performed the central calculations and analysis presented in this work. S.M. initially guided the analysis of this work. The manuscript was written by J.P. and D.M., and S.M. contributed to write the introduction and draw the conclusions of this manuscript.

%\section*{Data availability}
%The data that support the plots within this paper are available from the corresponding authors upon reasonable request.

%\section*{Competing interests}
%The authors declare no competing interests.

\bibliographystyle{apsrev4-2}
\bibliography{main}

\end{document}